\documentclass[a4paper,11pt]{article}
\usepackage{jheppub} % for details on the use of the package, please see the JINST-author-manual
\usepackage{lineno}
\usepackage{braket}
\usepackage{amsthm}
\newtheorem{theorem}{Theorem}[section]
\newtheorem{proposition}[theorem]{Proposition}
\newtheorem{lemma}[theorem]{Lemma}

\usepackage{xcolor}

\title{\boldmath Moore--Read construction and explicit monodromies of Laughlin states on Riemann surfaces}

\author{Kiyoon Eum}
\affiliation{Department of Mathematical Sciences, KAIST, 291 Daehak-ro, Yuseong-gu, Daejeon 34141, South Korea}

\emailAdd{kyeum@kaist.ac.kr}

\abstract{We revisit the Moore--Read construction of the $\nu=1/k$ Laughlin states on compact Riemann surfaces, deriving their conformal blocks from $U(1)_k$ Chern--Simons theory through the CS/WZW correspondence. We compute their explicit monodromies under quasi-hole transport and Aharonov--Bohm flux insertion, and prove that the monodromies associated with flux insertion asymptotically coincide with the corresponding physical adiabatic holonomies. In particular, we prove that the physical $L^2$ metric on the vector bundle of Laughlin states over the Jacobian is asymptotically projectively flat. We also relate our construction to the recently developed algebro-geometric approach to higher genus Laughlin states.
}

\begin{document}
\maketitle
\flushbottom

\section{Introduction}

In this paper, we revisit the Moore-Read construction of $\nu=1/k$ Laughlin states on Riemann surfaces. The starting point of \cite{moore1991nonabelions,read1992fractional} in construction of plane Laughlin states from conformal blocks was an effective $U(1)$ CS description of the FQHE system and CS/WZW duality \cite{witten1989quantum}. Then it was argued that the explicit monodromy of a conformal block coincides with the physical adiabatic holonomy of the quasi-hole positions in FQHE system, thus reproducing the famous fractional charge and fractional statistics of Laughlin quasi-holes \cite{Laughlin1983,ArovasSchriefferWilczek1984}. We will review this in Section \ref{section2}.  

As we move on to a Riemann surface $\Sigma$ of genus $g\geq 1$, we can consider new adiabatic transportations, namely transport of quasi-holes along non-trivial cycles on $\Sigma$ and the change of Aharonov--Bohm fluxes. The first part (Section \ref{section3}) of this paper is a Moore–Read construction of Laughlin states on Riemann surfaces, together with an explicit computation of their monodromies under quasi-hole transport and AB flux insertion. In fact, the Laughlin states on Riemann surfaces were already studied in many classical works and from many different perspectives, including \cite{HaldaneRezayi1985,Einarsson1990,Wen:1990zza,cristofano1991coulomb,cristofano1991hall,Li:1993tg,IengoLi1994,Alimohammadi:1998jh,spera2015geometry,Gromov:2016umy,Can:2017ycp}. Thus our result will recover those well-known results on Laughlin states on Riemann surfaces. Also, a free field construction of Laughlin states on higher genus Riemann surfaces was done in \cite{klevtsov2019laughlin}, which is slightly different from our construction since we start from CS/WZW correspondece. We intend this part of the paper to be semi-expository, organizing the various results on higher genus Laughlin states under the single viewpoint of Moore-Read construction.

Let us summarize the result of this part. Starting from the CS effective action, we use the abelian CS/WZW correspondence \cite{witten1989quantum,bos1989u,labastida1989chern,labastida1989operator} to obtain an explicit formula for the Laughlin states on Riemann surfaces with explicit dependence on external parameters. For the quasi-hole positions, it gives familiar fractional statistics and fractional AB phases. Along the non-trivial cycles it also gives the expected Heisenberg algebra \cite{Wen:1990zza}. For the AB fluxes, we obtain the rank $k^g$ bundle of Laughlin states $\mathcal{V}_k$ over $Jac(\Sigma)$, with slope $-1/k$. By a Niu--Thouless--Wu type formula this gives an averaged Hall conductance of $1/k$, as expected. The bundle $\mathcal{V}_k$ also comes with the natural theta metric whose Chern connection is projectively flat. Note that the projective flatness of the bundle of Laughlin states over $Jac(\Sigma)$ was suggested by \cite{Klevtsov:2021kii} as a test of the topological robustness of a quantum state of matter. However, the theta metric is not actually a physical inner product of quantum states.

In the second part of the paper, Section \ref{WenZeeSection}, we examine whether these formal results derived from the explicit formula are actually correct. First we introduce the Wen--Zee coupling to the CS action so that we have a correct relation between the number of the electrons and the total magnetic flux (Wen--Zee shift formula). After that, we can compare our construction with the algebro-geometric approaches to the Laughlin states in \cite{Klevtsov:2021kii,Burban:2023cii,klevtsov2025chern}; see Section \ref{4.2} for more about this. As a result, we identify our conformal block Laughlin states as axiomatic Laughlin state defined in \cite{klevtsov2019laughlin}, and our bundle of Laughlin states with the direct image Laughlin bundle constructed in \cite{klevtsov2025chern}. Then we prove that the physical $L^2$ metric is in fact asymptotically isometric to the natural theta metric on $\mathcal{V}_k$, as the number of electron tends to infinity. This shows that the formal derivation of Hall conductance using the theta metric in Section \ref{section3} is asymptotically correct.

\section{Review of Laughlin states and Moore--Read construction}\label{section2}

In \cite{moore1991nonabelions}, starting from an effective $U(1)_k$ Chern--Simons description of the abelian FQHE system, Moore and Read considered the $U(1)_k$ WZW model and showed that a correlator in the $U(1)_k$ WZW model with a background charge gives the $\nu=1/k$ Laughlin state. More precisely, consider the $U(1)_k$ WZW model with a free chiral field $\varphi$ satisfying $\varphi \sim \varphi+2\pi \sqrt{k}$. Denote the current operator and vertex operator by 
\begin{equation}\label{JV}
    J(z)=i\sqrt{k}\,\partial \varphi(z),\qquad V_\alpha(z)= :\exp{i\,\frac{\alpha}{\sqrt{k}}\,\varphi(z)}:.
\end{equation}
Then 
\begin{equation}\label{MooreRead}
 \Biggl\langle \prod_{a=1}^M V_{1}(\eta_a)\prod_{j=1}^N V_{k}(z_j) \exp{\frac{-i}{2\pi\sqrt{k}}\int d^2w \;\varphi(w)}\Biggl\rangle
 =\prod_{a<b}^M (\eta_a-\eta_b)^{1/k}\Psi_{L}(\eta;z)e^{-\frac{1}{4k}\sum_a |\eta_a|^2 },
\end{equation}
where $\Psi_{L}$ is a Laughlin state with $M$ quasi-holes and $N$ electrons in the symmetric gauge $A_z=-\bar{z}/4, A_{\bar{z}}=z/4$:
\begin{equation}\label{Laughlin}
 \Psi_{L}(\eta;z)=\prod_{a,j}(\eta_a-z_j)\prod_{i<j}^N(z_i-z_j)^k e^{-\frac{1}{4}\sum_j |z_j|^2 }.
\end{equation}
In fact, the exponential factor needs more care, and we will return to it shortly. 

The multi-valued holomorphic factor $\prod_{a<b}^M (\eta_a-\eta_b)^{1/k}$ in front of the expression \eqref{MooreRead} has the explicit monodromy of an anyon with fractional statistics $\theta/\pi=1/k$ as a quasi-hole coordinate $\eta_a$ goes around other quasi-hole coordinates. This is natural from the $U(1)_k$ Chern--Simons viewpoint where the insertion of the quasi-hole vertex operator $V_{1}$ corresponds to the insertion of a Wilson line operator, which describe a low energy excitation (quasi-holes) of the FQHE ground state. However, more is true. If we compute the norm of the state \eqref{Laughlin} and compute the Berry connection from it, the microscopic adiabatic transport gives the same fractional statistics. By the plasma analogy (see, for example,  \cite{stone1992quantum,tong2016lectures}), the norm of the Laughlin state \eqref{Laughlin} can be computed as 
\begin{equation}\label{plasma}
Z= \braket{\Psi_L|\Psi_L}\approx C \exp{\left(-\frac{1}{k}\sum_{a<b} \log |\eta_a-\eta_b|^2+\frac{1}{2k}\sum_a |\eta_a|^2 \right)},
\end{equation}
where the constant $C$ does not depend on the quasi-hole coordinates $\eta_a$.

The Berry connection with respect to quasi-hole coordinates $\eta_a$ is then computed as
\begin{equation}\label{Berry}
\mathcal{A}_{\eta_a}=\frac{1}{2}\frac{\partial \log Z}{\partial \eta_a}=-\frac{1}{2k}\sum_{a\neq b}\frac{1}{\eta_a-\eta_b}+\frac{\overline{\eta}_a}{4k},
\end{equation}
and
\begin{equation}\label{Berryanti}
\mathcal{A}_{\overline{\eta}_a}=-\frac{1}{2}\frac{\partial \log Z}{\partial \overline{\eta}_a}=\frac{1}{2k}\sum_{a\neq b}\frac{1}{\overline{\eta}_a-\overline{\eta}_b}-\frac{\eta_a}{4k}.
\end{equation}

As a quasi-hole coordinate $\eta_a$ moves along a closed path $C$ adiabatically, $\Psi_L$ acquires the phase
\begin{equation}\label{Berryphase}
    e^{i\gamma_{phase}}=\exp{\left(-\oint_C \mathcal{A}_{\eta_a}d\eta_a + \mathcal{A}_{\overline{\eta}_a}d\overline{\eta}_a \right)}=\exp{i\left( \gamma_{stat}-\frac{\Phi}{k} \right)},
\end{equation}
where $\Phi=\int_D \frac{i}{2}d\eta_a\wedge d\overline{\eta}_a$ represents the total background magnetic flux enclosed by the path $C=\partial D$ and $\gamma_{stat}=\frac{2\pi}{k}\times(\# \;\,\text{of} \,\; \eta_{b\neq a} \,\; \text{in} \;\, D)$. From this it follows that the quasi-holes have fractional charge $-1/k$ and fractional statistics parameter $1/k$.

This phenomenon, that is, the monodromy of the conformal block \eqref{MooreRead} coincides with the parallel transport with respect to the Berry connection, was the starting point for representing the holomorphic wave functions as conformal blocks in \cite{moore1991nonabelions}. This point was further examined in \cite{blok1992many,nayak19962n,read2009super} and its generalization is referred to as the \emph{Moore--Read conjecture} in \cite{hansson2017quantum}. Said differently, in this case the first part of the Berry connection (first term in \eqref{Berry}) coincides with the KZ connection \cite{knizhnik1984current} (the connection $1$-form in \eqref{Berry} is written in terms of a normalized state, hence the factor of $1/2$ is correct; see Appendix \ref{appen}). 

The topological part of \eqref{Berry} and \eqref{Berryphase} is now understood. What about the second term? It would be nice if the conformal block in \eqref{MooreRead} could give the Berry phase in the second term of \eqref{Berryphase} through its explicit monodromy. Unfortunately, this is not possible without a caveat. Let's see what can go wrong. First, observe that the holomorphic expression $\prod_{a<b}^M (\eta_a-\eta_b)^{1/k}$ satisfies the differential equation
\begin{equation}\label{diffeqn1}
    \left( \partial_{\eta_a} -\frac{1}{k}\sum_{b\neq a} \frac{1}{\eta_a-\eta_b} \right)\prod_{a<b}^M (\eta_a-\eta_b)^{1/k}=0,
\end{equation}
which is of course the KZ equation, and we just showed that this equals the parallel transport equation with respect to the first part of the Berry connection. If there exists a holomorphic expression $F(\eta_a)$ whose monodromy gives the second part of the Berry phase, it should satisfy the differential equation
\begin{equation}\label{diffeqn}
    \left( \partial_{\eta_a} +\frac{\overline{\eta_a}}{2k} \right)F(\eta_a)=0,
\end{equation}
which seems impossible since $\overline{\eta}_a$ is anti-holomorphic. However, the conformal block expression in fact provides a solution to \eqref{diffeqn} as much as it can. To understand this, we need to look more closely at the exponential factor in \eqref{MooreRead}. The Wick contraction between the vertex operator $V_{1}(\eta_a)$ and the background charge term $\exp{\frac{-i}{2\pi\sqrt{k}}\int d^2w \;\varphi(w)}$ produces an exponential term of the form
\begin{equation}\label{log}
F(\eta_a):= \exp{ \frac{-1}{2\pi k}\int d^2w \log(\eta_a-w) }.
\end{equation}

This multi-valued holomorphic expression is highly singular, since $\log(\eta_a-w)$ has a branch cut at every $w$ inside of the integral. In \cite{moore1991nonabelions} and \eqref{MooreRead}, only the real part of \eqref{log} is taken and yields a non-holomorphic Gaussian factor in the Laughlin state. However, rather than its value, one can consider how it changes as $\eta_a$ moves along a contour $C$. The phase change is then given by 
\begin{equation*}
\exp{\left( \frac{-i}{k}\text{Area}(D)\right)}=\exp{\left(-\frac{i\Phi}{k}\right)},
\end{equation*}
which is exactly the second term in \eqref{Berryphase} (see \cite{read1992fractional}). The singular expression \eqref{log} can be understood as follows. Instead of a continuous background charge density $1$, place normalized Dirac delta masses on a uniform grid so that the resulting discrete distribution converges to the uniform continuous charge distribution as the spacing $\propto l$ goes to zero. As long as $C$ does not pass through any grid point, the expression is well-defined and the phase change is then given by 
\begin{equation*}
\exp{\frac{-i l^2}{k}\left( \# \,\;\text{of grid points in}\,\; D\right)}\xrightarrow[l \to 0]{}\exp{\left( \frac{-i}{k}\text{Area}(D)\right)}.
\end{equation*}
Same consideration appeared in \cite{semenoff1989semenoff} and was also used in \cite{bradlyn2015topological}. We will see how this regularization process should be modified on Riemann surfaces.

We can also check that \eqref{log} indeed satisfies the equation \eqref{diffeqn}:
\begin{align*}
\partial_{\eta_a}F(\eta_a)&=\frac{-1}{2\pi k}\int d^2w\, \frac{1}{\eta_a-w}\,F(\eta_a)=\frac{-1}{2\pi k}\int d^2w\, \partial_{\overline{w}}\overline{w}\,\frac{1}{\eta_a-w}\,F(\eta_a)\\
&=\frac{1}{2\pi k}\int d^2w\, \overline{w}\,\partial_{\overline{w}}\frac{1}{\eta_a-w}\,F(\eta_a)=\frac{-1}{2\pi k}\int d^2w\, \overline{w}\,\pi \delta_{\eta_a}\,F(\eta_a)\\
&=-\frac{\overline{\eta}_a }{2k}F(\eta_a).
\end{align*}
Thus we found the holomorphic (although singular) solution to the non-holomorphic differential equation \eqref{diffeqn} via conformal block.

As in the case of the first term in \eqref{Berryphase}, this can be seen more clearly from the dual $U(1)_k$ Chern--Simons theory, now taking into account the background gauge field $A$. We will state the CS/WZW correspondence with background gauge field in question more precisely in the next section. For now, assume the correct CS Lagrangian is 
\begin{equation}\label{effective}
 S[a;A]=\frac{k}{4\pi}\int a\wedge da +\frac{1}{2\pi}\int a\wedge dA.
\end{equation}
This is also exactly the effective CS Lagrangian for the $\nu=1/k$ Laughlin states \cite{wen1992shift}. Here $a$ is dynamical and $A$ is a fixed background $U(1)$ gauge field. Completing the square, we can rewrite it as
\begin{equation*}
 S[a;A]=\frac{k}{4\pi}\int \left(a+\frac{A}{k}\right)\wedge d\left(a+\frac{A}{k}\right) -\frac{1}{4\pi k} \int A\wedge dA.
\end{equation*}
This is just the pure $U(1)_k$ CS Lagrangian for the new variable $\widetilde{a}= a+A/k$, and the corresponding Wilson line operators
\begin{equation}\label{Wilson}
    \widetilde{W}_n(\gamma)=\exp \left(in\int_\gamma \widetilde{a} \right)
\end{equation}
in time direction have statistical parameter $\theta/\pi=n^2/k$. Hence the Wilson line operators for $a$,
\begin{equation*}
    W_1(\gamma)=\exp \left(i\int_\gamma a \right)=\exp \left(i\int_\gamma \widetilde{a}-\frac{A}{k} \right)
\end{equation*}
obtain an additional phase
\begin{equation*}
    -\frac{i}{k}\oint_C A= -\frac{i}{k}\int_D dA =  -\frac{i}{k}\int_D F=-\frac{i}{k}\Phi
\end{equation*}
moving along a contour $C$, together with statistical phase coming from \eqref{Wilson}.\footnote{Whenever a $U(1)$ gauge field $a$ (or $A$) is written on the $2D$ spatial surface, we work in the $a_0=0$ gauge and again denote its spatial component by $a$.} Thus we recover \eqref{Berryphase} from the CS side, and the corresponding conformal block monodromy simply reflects this.

Let's summarize what we have seen so far. Starting from (effective) CS theory \eqref{effective}, we interpret a conformal block of the dual WZW model as a Laughlin state including some additional expressions for external parameters \eqref{MooreRead}, in this case the quasi-hole coordinates. Then the Berry phase coincides with the phase obtained from the CS theory, and this is explicitly reflected in the form of the conformal block. That is, the differential equations \eqref{diffeqn1}, \eqref{diffeqn} satisfied by the conformal block give directly the Berry connection \eqref{Berry}. Of course, this is only justified after computing the Berry connection via evaluating the norm of the state \eqref{plasma}. In any case, this is quite a remarkable coincidence, and it is natural to ask whether a similar coincidence can happen for other types of external parameters. One natural candidate for such an external parameter is the background gauge field $A$, and in this case the Berry curvature is closely related to the Hall conductance \cite{niu1985quantized}. To see this aspect of the story, we need to put Laughlin states on topologically non-trivial surfaces, that is, on higher genus Riemann surfaces. This is the topic of the next section.

\section{Moore--Read construction on higher genus Riemann surfaces}\label{section3}

\subsection{Moore--Read construction of Laughlin states}
Let $\Sigma$ be a compact Riemann surface of genus $g\geq 1$. The CS/WZW correspondence \cite{witten1989quantum, bos1989u, labastida1989operator, labastida1989chern} identifies the wavefunctional $\Psi_r[a_{\bar{z}}]$ of $U(1)_k$ CS theory 
\begin{equation*}
    S_{CS}[a]=\frac{k}{4\pi}\int_M a\wedge da
\end{equation*}
on $M=\mathbb{R}\times\Sigma$ with a $c=1$ conformal block of a compactified free scalar field $\varphi$ coupled to the background field $a$:
\begin{equation}\label{CS/WZW}
    \Psi_r[a^{0,1}]= \Biggl\langle \prod_{j=1}^N V_{\alpha_j}(z_j) \exp{\frac{-i}{2 \pi }\int_\Sigma\;a^{0,1}\wedge J}\Biggl\rangle_r,
\end{equation}
where the insertion of the vertex operator $V_{\alpha}(z)$ \eqref{JV} on the right-hand side corresponds to the insertion of the Wilson line source $W_{\alpha}(\gamma)$ \eqref{Wilson} piercing $\Sigma$ at $z$ on the left-hand side. Here  we normalized the chiral compact boson by $\varphi \sim \varphi+2\pi \sqrt{k}$ and $r\in (\mathbb{Z}/k\mathbb{Z})^g$ labels a conformal block.

We start with the effective CS Lagrangian with the background gauge field $A$:
\begin{equation*}
 S[a;A]=\frac{k}{4\pi}\int_M a\wedge da +\frac{1}{2\pi}\int_M a\wedge dA.
\end{equation*}
The background field $A$ is fixed and represents the strong background magnetic field in the quantum Hall system. Assume that $A$ is a connection form of the magnetic line bundle $L_\Phi$ of degree $N_\Phi \gg 0$. Completing the square, we get
\begin{equation}\label{CSlag}
 S[a;A]=S_{CS}[\widetilde{a}] -\frac{1}{4\pi k} \int_M A\wedge dA,
\end{equation}
with the new variable $\widetilde{a}= a+A/k$.\footnote{For the mathematical reader, assume $L_\Phi$ has a $k^{th}$ root.} The second term in \eqref{CSlag} gives the Hall conductance $1/k$ by the usual effective theory argument, but it does not play any role in what follows. Applying \eqref{CS/WZW} to \eqref{CSlag}, we have
\begin{equation}\label{MRonRS}
    \Psi_r[\widetilde{a}^{0,1}]= \Biggl\langle \prod_{a=1}^M V_{1}(\eta_a)\prod_{j=1}^N V_{k}(z_j) \exp{\frac{-i}{2 \pi }\int_\Sigma\;(a^{0,1}+\frac{A^{0,1}}{k})\wedge J}\Biggl\rangle_r.
\end{equation}
Note that we are only looking at the chiral half of the full CFT with $\phi(z,\bar{z})=\varphi(z)+\overline{\varphi}(\bar{z})$, and terms like $\int_\Sigma A^{0,1}\wedge \partial \varphi$ should be understood in the full CFT sense. This implies that the integration by parts should be done in the following way:
\begin{equation}\label{ibp}
  \int_\Sigma A^{0,1}\wedge \partial \varphi + A^{1,0}\wedge \bar{\partial} \overline{\varphi}=\int_\Sigma A\wedge d \phi = \int_\Sigma F\,\phi=\int_\Sigma F \varphi +F \overline{\varphi}.  
\end{equation}
Although $A$ and $\phi$ are multi-valued, this prescription of integration by parts does yield the correct result in the end once the smooth curvature integral is replaced by atomic masses of appropriate divisors. In particular, the coefficient in front of $\varphi$ becomes an integral multiple of $1/\sqrt{k}$ (see \cite[Section 4]{klevtsov2019laughlin} and the discussion at the end of this subsection). The rationale behind this is that we want to follow the plane Moore-Read construction as much as possible, which starts from the left-hand side of \eqref{MooreRead}. As a final check, if we put $a=0, A_{\bar{z}}=z/4$, we recover the background charge term of \eqref{MooreRead} in the plane case where $iF=d^2z$. Thus we expect the expression \eqref{MRonRS} to represent the explicit monodromy properties of the Laughlin state on Riemann surfaces. 

Before computing the correlator \eqref{MRonRS}, let us comment on the number of electrons $N$ and the variable $a_{\bar{z}}$. For simplicity, let $M=0$. Then the Gauss law constraint on the CS wave functional requires the condition
\begin{equation}\label{Gauss}
    \frac{k}{2\pi i}f+\frac{1}{2\pi i }F+k\sum_{j=1}^N \delta^{(2)}(z-z_j)=0,
\end{equation}
where $f=da, F=dA$. The degree of the line bundle $L_\Phi$ is given by $N_\Phi=\frac{i}{2\pi}\int_\Sigma F$. Let $N=N_\Phi/k$ and consider the situation where $N,N_\Phi \gg 0$. It is natural to assume that $N$ electrons (in this case represented by Wilson lines with charge $k$) usually tend to spread evenly across $\Sigma$ so that they screen the background magnetic field $F$ and satisfy $k\sum_{j=1}^N \delta^{(2)}(z-z_j)\approx \frac{i}{2\pi}F$ (cf. \eqref{plasma}). Then the resulting equation for $a$ becomes $f=0$, and after reduction, we assume that $a_{\bar{z}}$ belongs to $Jac(\Sigma)$ as in pure CS theory with no Wilson lines. 

Note however that this is mainly for convenience and does not play an important role in the main argument of this paper. One can always absorb the non-flat part of $a$ into $A$, especially on the CFT side, where both are fixed background fields. Nonetheless, it is nice to have $\int_\Sigma f=0$ at least to avoid confusion later. From now on, in this section we always assume $N+M/k=N_\Phi/k$ and $a_{\bar{z}}\in Jac(\Sigma)$. We will also see the same condition from charge neutrality on the CFT side. Note that this is different from the result of \cite{wen1992shift}; we will come back to this matter in Section \ref{WenZeeSection}. 

Now let us compute the expression \eqref{MRonRS}, following \cite{Eguchi:1986ui, dijkgraaf1988c, Verlinde:1986kw,Alvarez-Gaume:1987wwg, klevtsov2019laughlin}. We need to fix some notations. Let $\Sigma$ be a compact Riemann surface of genus $g\geq 1$, equipped with a canonical homology basis $(a_j, b_j)_{j=1,...,g}$. Let $\omega=(\omega_1,...,\omega_g)^T$ be the normalized basis of holomorphic one-forms, $\oint_{a_i}\omega_j=\delta_{ij}, \oint_{b_i}\omega_j=\tau_{ij}$ with period matrix $\tau$. Let $\Lambda=m+\tau n,\;m,n\in\mathbb{Z}^g$ be the lattice generated by the $g\times 2g$ matrix $(I,\tau)$. Then $Jac(\Sigma)=\mathbb{C}^g/\Lambda$ is called the Jacobian variety of $\Sigma$, and parametrizes the topologically trivial line bundles over $\Sigma$, which represent AB fluxes. The Abel-Jacobi map with the base point $z_0\in\Sigma$ is denoted by 
\begin{equation*}
    \mathcal{I}(z)=\int^z_{z_0}\omega \in Jac(\Sigma), \quad z\in\Sigma.
\end{equation*}
The level $k$ theta functions for $u\in\mathbb
{C}^g, r\in (\mathbb{Z}/k\mathbb{Z})^g$ are denoted by
\begin{align*}
    \Theta^{(k)}_r(u\mid\tau)&=\sum_{n\in\mathbb{Z}^g}\exp \left(\pi i k \left(n+\frac{r}{k}\right)^T \tau\left(n+\frac{r}{k}\right)+2\pi i k \left(n+\frac{r}{k}\right)^T u  \right)\\
    &=\theta\begin{bmatrix}r/k\\0\end{bmatrix}(ku\mid k\tau).
\end{align*}
The prime form is denoted by $E(z,w)$. Finally, let $\zeta_a$ be the $Jac(\Sigma)$ coordinates of $a_{\bar{z}}$, that is,
\begin{equation}\label{zeta}
    a^{0,1}=\bar{\partial}\chi+\pi \zeta_a^T(\mathrm{Im}\, \tau)^{-1}\overline{\omega},\quad \zeta_a=\frac{1}{2\pi i}\int_\Sigma a^{0,1}\wedge \omega.
\end{equation}
Under a large gauge transformation $a\longmapsto a+g^{-1}dg$ with winding numbers 
\begin{equation*}
    \int_{a_j}g^{-1}dg=2\pi i m_j, \quad \int_{b_j}g^{-1}dg=2\pi i n_j, 
\end{equation*}
 $\zeta_a\longmapsto \zeta_a +\tau m- n$ by the Riemann bilinear relations.

Now we can express \eqref{MRonRS} as
\begin{align}\label{formula}
 \mathcal F_r(\eta,z;a,A)
 =\,& \mathcal{N}(F,\tau)\,
 \Theta_r^{(k)}(U_{\widetilde{a}}\mid\tau)\prod_{a<b}E(\eta_a,\eta_b)^{1/k}\prod_{b,j}E(\eta_b,z_j)\prod_{i<j}E(z_i,z_j)^k \notag\\
 &\times\exp\left[-\frac{1}{k}\sum_b\int_\Sigma\log E(\eta_b,w)\,\frac{iF}{2\pi}(w)\right]\notag\\
 &\times\exp\left[-\sum_j\int_\Sigma\log E(z_j,w)\,\frac{iF}{2\pi}(w)\right],
\end{align}
where 
\begin{equation}\label{Ua}
    U_{\widetilde{a}}=\zeta_a+\frac{1}{k}\sum_b^M \mathcal{I}(\eta_b)+\sum_j^N \mathcal{I}(z_j)-\frac{1}{k}\int_\Sigma \mathcal{I}(w)\frac{iF}{2\pi}(w).
\end{equation}

To derive \eqref{formula}, we decompose $\varphi$ into a single-valued oscillatory part and an instanton part: $\varphi=\varphi_{\mathrm{osc}}+\varphi_{\mathrm{ins}}$. The oscillatory part is computed by Wick contractions using 
\begin{equation*}
    \langle\varphi_{\mathrm{osc}}(z)\varphi_{\mathrm{osc}}(w)\rangle
    =-\log E(z,w).
\end{equation*}
There are three types of contractions: $V-V$, $V-J$ and $J-J$ ($V$ denotes a vertex operator and $J$ the current operator). The $V-V$ contraction gives prime form parts in \eqref{formula}. The $V-J$ contraction is computed using \eqref{ibp}:
\begin{align}\label{intbypart}
 \exp \frac{1}{2 \pi }\int_\Sigma \widetilde{a}^{0,1}(w) i\partial_w \langle \varphi_{\mathrm{osc}}(\eta_a) \varphi_{\mathrm{osc}}(w)\rangle&=\exp \frac{-1}{2 \pi }\int_\Sigma i\partial_w\widetilde{a}^{0,1}(w) \log E(\eta_a,w)\notag\\
 &=\exp \frac{-1}{k}\int_\Sigma \log E(\eta_a,w) \frac{i}{2\pi}F(w). 
\end{align}
The $J-J$ contraction can be computed similarly to give a term of the form
\begin{equation*}
     \exp\left[\frac{1}{2k}\iint_{\Sigma\times\Sigma} \log E(w,w')\frac{iF}{2\pi}(w)\frac{iF}{2\pi}(w') \right],
\end{equation*}
and since it does not depend on $\eta_b, z_j, a$, we absorb it into the normalization constant $\mathcal{N}$. Note that the constant mode integral gives an integrated version of \eqref{Gauss}. 

Next we turn to the $\varphi_{\mathrm{ins}}$ part. Denote the chiral momentum lattice by $p=n+r/k, n\in \mathbb{Z}^g, r\in(\mathbb{Z}/k\mathbb{Z})^g$. Let 
\begin{equation*}
    \partial\varphi_{\mathrm{ins}}^{(p)}=2\pi \sqrt{k}p^T\omega, \quad \varphi_{\mathrm{ins}}^{(p)}(z)=\varphi_0+ 2\pi \sqrt{k} p^T \mathcal{I}(z).
\end{equation*}
We compute the fixed $p$ contribution. The propagator gives $\exp \left( \pi i k p^T \tau p\right)$ \cite{Eguchi:1986ui,dijkgraaf1988c}, and the vertex operators give
\begin{equation*}
    \exp i\frac{\alpha}{\sqrt{k}}\varphi_{\mathrm{ins}}^{(p)}(z)=\exp\left( 2\pi i\alpha p^T \mathcal{I}(z)\right).
\end{equation*}
The background magnetic field term similarly gives 
\begin{equation*}
    \exp\frac{-i}{2 \pi }\int_\Sigma\;\frac{A^{0,1}}{k}\wedge J_{\mathrm{ins}}^{(p)}=\exp \left(-2\pi ip^T\int_\Sigma \mathcal{I}(w)\frac{iF}{2\pi}(w)\right).
\end{equation*}
Finally, the $a^{0,1}$ term gives
\begin{align*}
    \exp\frac{-i}{2\pi }\int_\Sigma\;a^{0,1}\wedge J_{\mathrm{ins}}^{(p)}&=\exp \frac{-i}{2\pi }\int_\Sigma \pi \zeta_a^T (\mathrm{Im}\,\tau)^{-1}\overline{\omega} \wedge i 2\pi k \, p^T \omega\\
    &=\exp \pi k(\zeta_a)_i (\mathrm{Im}\,\tau)^{-1}_{ij} p_l\int_\Sigma \overline{\omega}_j \wedge \omega_l \\
    &=\exp \pi k(\zeta_a)_i (\mathrm{Im}\,\tau)^{-1}_{ij} p_l(2i\mathrm{Im}\,\tau)_{jl}\\
    &=\exp \left( 2\pi i k \, p^T\zeta_a\right).
\end{align*}
Combining these and summing over $n\in\mathbb{Z}^g$, we get (recall the definition of $U_{\widetilde{a}}$ \eqref{Ua})
\begin{equation*}
   \sum_{n\in\mathbb{Z}^g}\exp\left( \pi i k p^T \tau p+2\pi i k p^T U_{\widetilde{a}}\right)=\Theta_r^{(k)}(U_{\widetilde{a}}\mid\tau).
\end{equation*}
This completes the computation of \eqref{formula}.

Singular expressions such as $\int_\Sigma \log E(z,w)\frac{iF}{2\pi}(w)$ can be dealt with as in the plane case, by replacing $iF/2\pi$ with atomic masses. However, on Riemann surfaces $\varphi$ is multi-valued and the operator $e^{i\alpha/\sqrt{k}\, \varphi}$ is not well-defined unless $\alpha$ is integral. Thus we must use a unit mass magnetic divisor $D_m=\sum^{N_\Phi}_{\mu=1} w_\mu$ such that $L_\Phi \simeq \mathcal{O}(D_m)$ (which is the curvature of the singular metric defined by a global holomorphic section; see \cite[Section 4.4]{klevtsov2019laughlin}), and instead of small grid length limit $l\rightarrow 0$, we need to take the $N_\Phi \rightarrow \infty$ limit. As a result we get expressions such as
\begin{equation*}
    \prod_{b,\mu} E(\eta_b,w_\mu)^{-1/k}\prod_{j,\mu} E(z_j,w_\mu)^{-1},
\end{equation*}
and
\begin{equation}\label{Ua2}
    U_{\widetilde{a}}=\zeta_a+\frac{1}{k}\sum_b^M \mathcal{I}(\eta_b)+\sum_j^N \mathcal{I}(z_j)-\frac{1}{k}\mathcal{I}(D_m).
\end{equation}
Note that \eqref{Ua2} is independent of the base point of the Abel-Jacobi map only if $N+M/k=N_\Phi/k$.

\subsection{Monodromy property for quasi-holes}\label{qhsubsection}

For a contractible curve $C=\partial D$ on $\Sigma$, it is clear from \eqref{formula} that a quasi-hole coordinate has the same monodromy phase \eqref{Berryphase} as in the plane case. We now compute directly from \eqref{formula} what happens when one quasi-hole is transported around a non-trivial cycle. 

Fix a quasi-hole coordinate $\eta_c$, and let $e_\ell=(0_1,...,1_\ell,...,0_g)^T\in \mathbb{Z}^g$. The transport of $\eta_c$ around the canonical cycles gives
\begin{equation}\label{AJcontinuation}
    \mathcal{I}(\eta_c) \longmapsto \mathcal{I}(\eta_c)+ e_\ell \quad \text{along }a_\ell,\qquad \mathcal{I}(\eta_c) \longmapsto \mathcal{I}(\eta_c)+\tau e_\ell \quad \text{along }b_\ell.
\end{equation}
In the standard prime form trivialization, its corresponding automorphy properties are
\begin{align}\label{primeautomorphy}
    E(z+a_\ell,w)&=E(z,w), \notag\\
    E(z+b_\ell,w)&=\exp\left[-\pi i\tau_{\ell\ell}-2\pi i\left(\mathcal{I}_\ell(z)-\mathcal I_\ell(w)\right)\right] E(z,w).
\end{align}
Here $z+a_\ell$ and $z+b_\ell$ denote the denote analytic continuation along the chosen loops $a_\ell, b_\ell$. Strictly speaking, the prime form is a $(-1/2,-1/2)$-form, and we suppressed the half-form factors in \eqref{primeautomorphy} for simplicity; see \cite[(1.4)]{Fay1992KernelFA}. To fully describe the monodromies of quasi-holes on Riemann surfaces, including both local and global monodromies, we in fact need to consider those half-form factors as well. The result will be representation of the surface \textit{framed} braids group; see \cite{Li:1993tg,Bergeron:1993rb}. In the following we compute the algebra \eqref{algebra} they satisfy. Since these factors are common to all block labels $r$, their commutation relations are independent of the scalar normalization.

We also need the quasi-periodicity of the level $k$ theta functions:
\begin{align}\label{quasiperTheta}
    \Theta_r^{(k)}\left( u+\frac{n}{k}\,\middle|\, \tau \right) &=\exp \left(\frac{2\pi i}{k} r^T n\right)\Theta_r^{(k)}(u\mid\tau), \notag\\
    \Theta_r^{(k)}\left( u+\frac{\tau m}{k}\,\middle|\, \tau \right) &=\exp \left(-\frac{\pi i}{k} m^T\tau m -2 \pi i m^Tu \right) \Theta_{r+m}^{(k)}(u\mid\tau).
\end{align}
All conformal block labels $r$ below are understood modulo $k$.

First, transport $\eta_c$ around $a_\ell$. Equations \eqref{Ua} and
\eqref{AJcontinuation} give
\begin{equation*}
    U_{\widetilde a} \longmapsto U_{\widetilde a}+\frac{e_\ell}{k}.
\end{equation*}
The first identity in \eqref{quasiperTheta} therefore multiplies the theta factor by $e^{2\pi i r_\ell/k}$. All prime form factors involving $\eta_c$ are invariant under $z\longmapsto z+a_\ell$. Thus the $a_\ell$-cycle monodromy is
\begin{equation}\label{Ma}
    \mathsf{M}_{a_\ell}\mathcal F_r = e^{2\pi i r_\ell/k}\mathcal F_r.
\end{equation}

The $b_\ell$-cycle case is more interesting, since every factor in
\eqref{formula} involving $\eta_c$ contributes. Since
\begin{equation*}
    U_{\widetilde a}\longmapsto U_{\widetilde a}+\frac{\tau e_\ell}{k},
\end{equation*}
the theta factor transforms as
\begin{equation}\label{btheta}
    \Theta_r^{(k)}(U_{\widetilde a}\mid\tau) \longmapsto e^{-\pi i\tau_{\ell\ell}/k-2\pi i(U_{\widetilde a})_\ell} \Theta_{r+ e_\ell}^{(k)}(U_{\widetilde a}\mid\tau).
\end{equation}
On the other hand, the prime forms in quasi-hole--quasi-hole factors, the quasi-hole--electron factors, and the quasi-hole--background factor respectively acquire the phases
\begin{align*}
\mathcal{R}_{\ell}^{qq} &=\exp\left[ -\frac{\pi i(M-1)}{k}\tau_{\ell\ell} -\frac{2\pi i}{k}\left((M-1)\, \mathcal{I}_\ell(\eta_c)-\sum_{d\neq c} \mathcal{I}_\ell(\eta_d) \right) \right], \\
\mathcal{R}_{\ell}^{qe} &=\exp\left[ -\pi iN\tau_{\ell\ell} -2\pi i\left(N \,\mathcal{I}_\ell(\eta_c)-\sum_{j=1}^N \mathcal{I}_\ell(z_j) \right) \right], \\
\mathcal{R}_{\ell}^{F} &= \exp\left[ \frac{\pi iN_\Phi}{k}\tau_{\ell\ell} +\frac{2\pi i}{k}\left(N_\Phi\, \mathcal{I}_\ell(\eta_c) - \int_\Sigma\mathcal I_\ell(w)\,\frac{iF}{2\pi}(w) \right) \right].
\end{align*}
The charge neutrality condition $N+M/k=N_\Phi/k$ implies
\begin{equation*}
    -\frac{M-1}{k}-N+\frac{N_\Phi}{k}=\frac{1}{k}.
\end{equation*}
Using this identity and the definition \eqref{Ua}, we obtain
\begin{equation}\label{R}
\mathcal{R}_{\ell}^{qq} \mathcal{R}_{\ell}^{qe} \mathcal{R}_{\ell}^{F} = \exp\left[ \frac{\pi i}{k}\tau_{\ell\ell} +2\pi i\left( (U_{\widetilde a})_\ell - (\zeta_a)_\ell \right) \right].
\end{equation}
Combining \eqref{R} and \eqref{btheta}, the $b_\ell$-cycle monodromy is
\begin{equation}\label{Mb}
    \mathsf{M}_{b_\ell}\mathcal F_r = e^{-2\pi i(\zeta_a)_\ell} \mathcal F_{r+ e_\ell}.
\end{equation}
Thus the operators $\mathsf{M}_{a_\ell}, \mathsf{M}_{b_\ell}$ form a well-known algebra \cite{Wen:1990zza} (cf. \cite{cristofano1991hall}):
\begin{equation}\label{algebra}
\mathsf{M}_{a_\ell}\mathsf{M}_{a_m} = \mathsf{M}_{a_m}\mathsf{M}_{a_\ell},\quad   
\mathsf{M}_{b_\ell}\mathsf{M}_{b_m} =\mathsf{M}_{b_m}\mathsf{M}_{b_\ell}, \quad  
\mathsf{M}_{a_\ell}\mathsf{M}_{b_m} = e^{2\pi i \delta_{\ell m}/k} \mathsf{M}_{b_m}\mathsf{M}_{a_\ell}.
\end{equation}
Note that \eqref{Ma}, \eqref{Mb} are independent of $\eta_c$. The analogous monodromy operators for electron coordinates (which are just $\mathsf{M}_{a_\ell}^k, \mathsf{M}_{b_\ell}^k$) also do not depend on $z_j$ and $r$. This hints that \eqref{formula} might be made into a holomorphic section of the same line bundle for each $z_j$, for all $r$. See Section \ref{WenZeeSection}.

\subsection{Monodromy property for AB fluxes}\label{ABsubsection}

Now we consider the monodromy of \eqref{formula} with respect to the flat part of the background magnetic connection $A$. That is, we change the flat part of the connection $A$, while keeping its curvature $F=dA$ fixed. As a result, we compute the Hall conductance, generalizing \cite{cristofano1991hall}. Fixing a reference connection $A_0$ with curvature $F$, $A$ may be written as
\begin{equation*}
    A=A_0+A_\xi, \quad dA_\xi=0.
\end{equation*}
Let $\xi=x+\tau y, \, x,y\in \mathbb{R}^g$ be the holomorphic $Jac(\Sigma)$-coordinates of $A_\xi$ as in \eqref{zeta}:
\begin{equation}\label{jaccoor}
    \xi=\frac{1}{2 \pi i}\int_\Sigma A_\xi^{0,1}\wedge \omega.
\end{equation}
Under a large gauge transformation of $A_\xi$, $\xi \longmapsto \xi + n +\tau m$. Since $\widetilde a=a+A/k$, the flat part of $A$ modifies the theta argument in \eqref{formula} by $\zeta_a \longmapsto \zeta_a+\xi/k$ (this will be derived more precisely in Section \ref{subsectionproof}). All the other factors in \eqref{formula} depend on $A$ only through $F$ and are therefore unchanged by the variation of $A_\xi$. Consequently, the entire $\xi$-monodromy is contained in the level $k$ theta function with coefficient $1/k$, and from now on we identify $\mathcal{F}_r(\cdot)$ with $\Theta_r^{(k)}(\cdot/k \mid \tau)$ in this subsection. That is, for now we ignore the $\zeta_a$ and $\eta,z$-dependent terms in $U_{\tilde{a}}$ and only retain the residual automorphy that defines the rank $k^g$ holomorphic theta bundle. 

From the quasi-periodicity properties of the level $k$ theta functions \eqref{quasiperTheta}, we have
\begin{equation}\label{factorofaut}
\mathcal{F}_r(\xi+n+\tau m)=\exp\left( \frac{2\pi i}{k}r^T n-\frac{\pi i}{k}m^T \tau m -\frac{2\pi i}{k} m^T \xi \right) \mathcal{F}_{r+m}(\xi).
\end{equation}
Under a $\tau$-period, these transformations permute the block labels, and hence a single $\mathcal{F}_r$ cannot form a line bundle over $Jac(\Sigma)$. Let 
\begin{equation*}
    \mathcal{F}(\xi)=(\mathcal{F}_r(\xi))_{r\in (\mathbb{Z}/k\mathbb{Z})^g}
\end{equation*}
be the column vector of $k^g$ blocks. Then \eqref{factorofaut} can be written as
\begin{equation*}
    \mathcal F(\xi+n+\tau m)=J_{n,m}(\xi)\mathcal F(\xi),
\end{equation*}
with the $\mathrm{GL}(k^g,\mathbb{C})$-valued factor of automorphy $J_{n,m}$. It defines a rank $k^g$ vector bundle $\mathcal{E}_k$ over $Jac(\Sigma)$.\footnote{See \cite{kobayashi2014differential} for the differential geometric background of what follows.} Details are provided in Appendix \ref{appenEk}. Here we compute its curvature. To do so, we need to fix the connection on $\mathcal{E}_k$. Define the natural metric (cf. \cite[(31)]{bos1989u}, \cite[(15)]{labastida1989operator} and \cite[(4.13)]{bradlyn2015topological}) by
\begin{equation}\label{H}
H(\xi)=\exp\left( -\frac{2\pi}{k}\,\mathrm{Im}\,\xi^T (\mathrm{Im}\,\tau)^{-1}\mathrm{Im}\,\xi \right)\mathbf{1}_{k^g}.
\end{equation}
It is easy to verify that $H$ gives a well-defined Hermitian metric on $\mathcal{E}_k$, as in \eqref{metricproof}. The standard formula for the curvature of the Chern connection gives the curvature form
\begin{align}\label{curvature}
F_{\nabla^{\mathcal{E}_k}}&=\bar{\partial}\left( H^{-1}\partial H \right)=\bar{\partial}\left(\frac{2\pi i}{k} \mathrm{Im}\,\xi^T (\mathrm{Im}\,\tau)^{-1} d\xi \right)\mathbf{1}_{k^g} \notag\\
&=\frac{\pi}{k}\, d\xi^T (\mathrm{Im}\,\tau)^{-1}\wedge d\bar{\xi}\;\mathbf{1}_{k^g}=-\frac{2\pi i}{k}\,\omega_{\mathrm{flux}} \;\mathbf{1}_{k^g},
\end{align}
where $\omega_{\mathrm{flux}}$ is the canonical K\"ahler form on $Jac(\Sigma)$,
\begin{equation*}
 \omega_{\mathrm{flux}}=\frac{i}{2}d\xi^T (\mathrm{Im}\,\tau)^{-1}\wedge d\bar{\xi}=dx^T\wedge dy.
\end{equation*}
Note that since the curvature form \eqref{curvature} is a scalar multiple of the identity, it follows that $\nabla^{\mathcal{E}_k}$ is a projectively flat connection and $\mathcal{E}_k$ is a projectively flat bundle \cite[Proposition 1.2.8]{kobayashi2014differential}. 

The first Chern form of $\mathcal{E}_k$ over $Jac(\Sigma)$ is therefore
\begin{equation*}
    c_1(\mathcal{E}_k,H)=\frac{i}{2\pi}\mathrm{Tr}(F_{\nabla^{\mathcal{E}_k}})=k^{g-1}\omega_{\mathrm{flux}}.
\end{equation*} With the Niu--Thouless--Wu formula \cite{niu1985quantized,Tao:1984vy,Avron85,Klevtsov:2021kii}, this yields the fractional Hall conductance
\begin{equation}\label{Hallcon}
    \sigma_H=\frac{1}{\mathrm{rk}\mathcal{E}_k}c_1(\mathcal{E}_k,H)=\frac{1}{k}\,\omega_{\mathrm{flux}}
\end{equation}
and the average version
\begin{equation}\label{Hallcon2}
    \overline{\sigma_H}=\frac{1}{g!}\int_{Jac(\Sigma)}\sigma_H \wedge \omega^{g-1}_{\mathrm{flux}}=\frac{1}{k}.
\end{equation}
To be more precise, the Niu--Thouless--Wu formula relates $\sigma_H$ to the curvature of the dual bundle $\mathcal{E}_k^*$ by $\sigma_H=-i\operatorname{Tr}F_{\nabla^{\mathcal{E}_k^*}}/(2\pi \,\mathrm{rk}\mathcal{E}_k^*)$; see the derivation in \cite[Section 2.2.4]{tong2016lectures}. The dual bundle $\mathcal{E}_k^*$ will be identified with the bundle of Laughlin states in Section \ref{4.2}. 

We would like to emphasize that \eqref{Hallcon} is not obtained from electromagnetic response of \eqref{CSlag}. We derive it from the monodromy property \eqref{factorofaut} of \eqref{formula} in $A_\xi$ and a Niu--Thouless--Wu type formula. In particular, \eqref{Hallcon2} is a cohomological statement and independent of the choice of $H$. Later in Subsection \ref{subsectionproof}, we will show that the physical $L^2$ metric is in fact almost identical to $H$, hence \eqref{Hallcon} is also almost correct.

\section{Wen--Zee shift and more rigorous approaches to Laughlin states}\label{WenZeeSection}

\subsection{Wen--Zee shift formula}
Assume $M=0$ for now. The neutrality condition $N=N_\Phi/k$ that appeared in the preceding section does not contain the shift term $\mathcal{S}$ in the Wen--Zee shift formula \cite{wen1992shift}. This inconsistency can be resolved if we add a Wen--Zee term to the CS Lagrangian \eqref{CSlag}. Indeed, we expect a wave function of a charged particle in a magnetic field to transform as a scalar under coordinate transformations and by phases under a gauge transformation. However, in the previous section the electron vertex operator $V_{k}$ has conformal dimension $k/2$, and therefore the conformal blocks constructed above describe particles of gravitational spin $s=k/2$.  We will see that the Wen--Zee coupling gives the standard curvature background charge on the CFT side and changes the conformal dimension to a prescribed $s$. This kind of derivation of the Wen--Zee shift from CFT charge neutrality already appeared in \cite{klevtsov2019laughlin}; we connect this to the CS side.

Let $\varpi$ be the Levi--Civita connection on $T^{1,0}\Sigma=K_\Sigma^{-1}$ and set $\mathcal{R}=d\varpi$. In this notation, the Gauss--Bonnet theorem reads
\begin{equation*}
    \frac{i}{2\pi}\int_\Sigma\mathcal{R} =\frac{1}{4\pi}\int_\Sigma R_g\,dV_g= \chi(\Sigma)=2-2g.
\end{equation*}

Following \cite{wen1992shift,bradlyn2015topological}, let $\overline{s}=k/2-s$ be the mean orbital spin per particle. We consider the modified CS Lagrangian with a Wen--Zee term
\begin{equation}\label{CSWZlag}
S_s[a;A,\varpi]=\frac{k}{4\pi}\int_M a \wedge da +\frac{1}{2\pi}\int_M a \wedge dA +\frac{\overline{s}}{2\pi}\int_M a\wedge d\varpi.
\end{equation}
Define
\begin{equation*}
A_s = A+\overline{s}\,\varpi,\quad F_s=dA_s,\quad \widetilde{a}_s =a+\frac{A_s}{k}.
\end{equation*}
Here $A_s$ is a connection on $L_\Phi\otimes K^{-\overline{s}}_\Sigma$. We henceforth assume $s,\overline{s}\in\frac{1}{2}\mathbb{Z}$ and fix a spin bundle $S_\delta$ when a half-integer power of $K_\Sigma$ occurs. Denote the vector of Riemann constants by $\Delta$.

Now we can repeat the procedure of the previous section with $\widetilde{a}$ replaced by $\widetilde{a}_s$. The special case $s=k/2$, $\overline{s}=0$ is the one considered in the previous section. Completing the square in \eqref{CSWZlag}, we get
\begin{equation*}
 S_s[a;A,\varpi]=S_{CS}[\widetilde a_s] -\frac{1}{4\pi k}\int_M A_s \wedge d A_s .
\end{equation*}
As before, we restrict to the $da=0$ sector, in which the Gauss-law constraint becomes
\begin{equation}\label{Sgauss}
k\sum_{j=1}^N \delta^{(2)}(z-z_j)=\frac{i}{2\pi}F_s=\frac{i}{2\pi}F+\frac{i}{2\pi}\overline{s}\,\mathcal{R}.
\end{equation}
Integrating \eqref{Sgauss} over $\Sigma$ and substituting $s=0$, we get
\begin{equation}\label{Sneutrality}
N=\frac{N_\Phi}{k}+\frac{\overline{s}}{k}(2-2g)=\frac{N_\Phi}{k}+1-g.
\end{equation}
Thus we recover the shift formula of Wen--Zee (with $t=1, s=k/2$ in their paper).

The corresponding conformal block now contains an additional term
\begin{equation}\label{Rcoupling}
    \exp{\frac{-i\,\overline{s}}{2 \pi }\int_\Sigma\;\frac{\varpi^{0,1}}{k}\wedge J}=\exp \frac{\overline{s}}{2 \pi \sqrt{k}}\int_\Sigma\;\mathcal{R}\,\varphi=\exp \frac{-i\,\overline{s}}{4\pi\sqrt{k}}\int_\Sigma \,\varphi\, R_g \,dV_g.
\end{equation}
Equivalently, the Euclidean action acquires the background charge term
\begin{equation*}
    \Delta S_{E}= \frac{i\widetilde{Q}}{8\pi}\int_\Sigma \,\varphi\, R_g \,dV_g,\qquad \widetilde{Q}=\frac{2\,\overline{s}}{\sqrt{k}},
\end{equation*}
and modifies the conformal dimension of the electron vertex operator $V_k$ to (see \cite{DiFrancesco:1997nk})
\begin{equation*}
    h=\frac{k}{2}\longmapsto h_s=\frac{\sqrt{k}}{2}\left( \sqrt{k} - \widetilde{Q}\right)=s.
\end{equation*}
Hence we get a zero conformal dimension for electron operators when $s=0$. Again, we can obtain the shift formula \eqref{Sneutrality} from the charge neutrality condition on the CFT side. Denote the resulting conformal block by $\mathcal{F}^s_r$ so that $\mathcal{F}_r=\mathcal{F}_r^{\,k/2}$ (recall \eqref{formula}).

\subsection{Algebro-geometric definition of Laughlin states on Riemann surfaces}\label{4.2}

We now compare the \emph{"Laughlin states"} $\mathcal{F}_r^s$ constructed above from conformal blocks with the algebro-geometric definition of Laughlin states introduced by Klevtsov \cite{klevtsov2019laughlin}. Its formulation in terms of line bundles on symmetric powers and the corresponding construction of vector bundles over Picard varieties were developed in \cite{Klevtsov:2021kii,klevtsov2025chern}. The genus-one multilayer case was studied in \cite{Burban:2023cii}, a higher genus multilayer extension was developed in \cite{aldonza2025chern}, and the case with localized quasi-holes was treated in \cite{dupont2026chern}. In particular, the physical Berry connection is explicitly computed in the genus $g=1$ case \cite{Burban_Klevtsov_2025}.

We follow the definition given in \cite{klevtsov2019laughlin, dupont2026chern}. See ibid. for more details. Let $\pi_j:\Sigma^N\to\Sigma$ be the projections. Let
\begin{equation*}
    \Delta_N=\sum_{i<j}^N \Delta_{ij}, \qquad W_\eta=\sum_{j=1}^N\sum_{a=1}^M \pi_j^{-1}(\eta_a)
\end{equation*}
be the big diagonal divisor and the quasi-hole divisor on $\Sigma^N$, respectively. Let $L_\Phi$ be the magnetic line bundle, whose holomorphic sections correspond to lowest Landau level electron wave functions. The space of Laughlin states with fixed quasi-holes at $(\eta_a)_{a=1,...,M}$ is defined to be
\begin{equation}
H^0\left( \Sigma^N,L_\Phi^{\boxtimes N} \otimes\mathcal{O}(-k\Delta_N- W_\eta) \right)^{\mathfrak{S}_N}.
\end{equation}
Here the superscript denotes the invariant subspace under the action of the permutation group $\mathfrak{S}_N$, acting anti-symmetrically if $k$ is odd. Equivalently, the Laughlin states $F$ are defined by the following three conditions:
\begin{enumerate}
    \item with all variables except $z_j$ fixed, $F$ is a holomorphic section of $L_\Phi$ in $z_j$;
    \item $F$ vanishes to order at least $k$ along every particle diagonal $\Delta_{ij}$, and to order at least one along $z_j=\eta_a$;
    \item $F$ is alternating for odd $k$ and symmetric for even $k$.
\end{enumerate}

With this definition, it has been rigorously proved \cite{klevtsov2025chern, dupont2026chern} that there are exactly $k^g$ linearly independent Laughlin states and the numbers of electrons and quasi-holes satisfy the Wen--Zee type formula
\begin{equation*}
    N_\Phi=kN+M+k(g-1).
\end{equation*}
This agrees with \eqref{Sneutrality} for $s=0$ and there are exactly $k^g$ conformal blocks as $r$ runs over $(\mathbb{Z}/k\mathbb{Z})^g$. Moreover, the characteristic classes of the bundle of Laughlin states over $Pic^0(\Sigma)\simeq Jac(\Sigma)$ have been computed, and they coincide with those obtained in Section \ref{ABsubsection}. Hence the very natural question is what the relation is between $\mathcal{F}^{\,0}_r$ and this axiomatic definition of Laughlin states. Below we address this question.

First, we must admit that we were imprecise in the computation of \eqref{MRonRS} and we cannot get $\mathcal
{F}_r^s$ by simply replacing $A$ with $A_s$. Since we wanted a version of the Moore-Read construction \eqref{MooreRead} on Riemann surfaces, we proceeded to treat the $iF/2\pi$ term as a distribution of screening charges compensating for the insertion of electron and quasi-hole vertex operators, and considered only an atomic regularized version of \eqref{formula}. Moreover, if we were to include a genuine $A$ coupling, it would break the conformal symmetry and we could not factor the chiral half of the WZW model.

To actually compute the $\mathcal{R}$ coupling term, we must proceed as in \cite{Verlinde:1986kw,Alvarez-Gaume:1987wwg,klevtsov2019laughlin}. 
In particular, setting $Q=-\sqrt{k}\widetilde{Q}=-2\,\overline{s}$ in \cite[(6.21)]{Verlinde:1986kw}, we obtain:
\begin{equation}\label{formula2}
 \mathcal F^s_r(z)
 = \mathcal{N}(w,\eta)\, \Theta_r^{(k)}(U\mid\tau)\prod_{b,j}E(\eta_b,z_j)\prod_{i<j}E(z_i,z_j)^k \prod_{j,\mu}E(z_j,w_\mu)^{-1}\prod_j \sigma(z_j)^{Q},
\end{equation}
where 
\begin{equation*}
    U=\frac{1}{k}\sum_b^M \mathcal{I}(\eta_b)+\sum_j^N \mathcal{I}(z_j)-\frac{1}{k}\mathcal{I}(D_m)+\frac{Q}{k}\Delta, \quad D_m=\sum_\mu^{N_\Phi} w_\mu,
\end{equation*}
and 
\begin{equation*}
\sigma(z) =\exp\left[ -\sum_{\ell=1}^g \oint_{a_\ell} \omega_\ell(x)\log E(x,z) \right]
\end{equation*}
is Fay's multiplicative $g/2$-differential \cite{Fay1992KernelFA}. Indeed, the half-form automorphy factors of the prime form and $\sigma$ ensure that \eqref{formula2} has conformal dimension $s$ by \eqref{Sneutrality}.

To absorb the poles in \eqref{formula2}, let
\begin{equation*}
    \rho(z)=\sigma(z)^{N_\Phi/g}\prod_\mu^{N_\Phi} E(z,w_\mu).
\end{equation*}
This is a holomorphic section of $\mathcal{O}(D_m)$. Finally, define
\begin{align}\label{klevformula}
    F^s_r(z)&:=\mathcal{F}_r^s(z) \times \prod_j^N \rho(z_j)\notag\\
    &=\mathcal{N}\,\Theta_r^{(k)}(U\mid\tau)\prod_{b,j}E(\eta_b,z_j)\prod_{i<j}E(z_i,z_j)^k \prod_j \sigma(z_j)^{\frac{1}{g}(kN+M+2s-k)}.
\end{align}
Setting $s=0, \overline{s}=k/2$, we recover exactly the explicit $k^g$ Laughlin states given in \cite[(3.29)]{klevtsov2019laughlin}, which satisfy the above axioms.

Now we compare the Laughlin bundle in \cite{klevtsov2025chern} with our automorphy bundle $\mathcal{E}_k$ in the previous section. Note that the automorphy bundle $\mathcal{E}_k$ is \emph{not} a bundle consisting of Laughlin states. For $N \gg 1$, the components of $F(\xi)=(F_r^{\,0}(\xi))_r$ form a basis for the space of Laughlin states, whereas $F$ itself is a section of $\mathcal{E}_k$ over $\xi\in Jac(\Sigma)$. Thus the fiber of $\mathcal{E}_k\vert_\xi$ is not the space of Laughlin states. Instead, any Laughlin state $\Psi_L$ is given by a linear combination of $F_r^{\,0}$:
\begin{equation*}
    \Psi_L(\xi)=c(\xi)^T F(\xi), \qquad c(\xi)\in \mathbb{C}^{k^g}.
\end{equation*}
For the Laughlin state $\Psi_L$ defined in this way to be well-defined, the coefficients $c(\xi)$ must transform under $\xi \longmapsto \xi+n+\tau m$ by
\begin{equation*}
    c(\xi+n+\tau m)=\left( J_{n,m}(\xi)^{-1}\right)^T c(\xi).
\end{equation*}
This is a factor of automorphy for the dual bundle of $\mathcal{E}_k$, and we obtain a vector bundle of Laughlin states $\mathcal{V}_k$ by setting
\begin{equation*}
    \mathcal{V}_k:=\mathcal{E}_k^*.
\end{equation*}
Since we already computed the characteristic classes of $\mathcal{E}_k$, we immediately get
\begin{equation*}
    \mathrm{ch}_i(\mathcal{V}_k)=(-1)^i\mathrm{ch}_i(\mathcal{E}_k)=k^{g-i}\frac{[-\omega_{\mathrm{flux}}]^i}{i!},\qquad \mathrm{ch}(\mathcal{V}_k)=k^{g}e^{-[\omega_{\mathrm{flux}}]/k}.
\end{equation*}
This coincides with \cite[Theorem 2, $p=0$]{klevtsov2025chern} and suggest that they might be holomorphically isomorphic bundles. Let $V_{N,d,b,g}$ be the Laughlin vector bundle constructed in \cite{klevtsov2025chern} with $0=p=d-b(N+g-1)$.\footnote{Identification $Jac(\Sigma)\simeq \mathrm{Pic}^d(\Sigma)$ is understood.} In our notation, $d=N_\Phi, b=k$. Then by the result of Mukai \cite[Proposition 6.17]{mukai1978semi}, if both $\mathcal{V}_k$ and $V_{N,d,b,g}$ are simple semi-homogeneous, they are isomorphic up to a flat twist. We show that $\mathcal{V}_k$ is simple semi-homogeneous in Appendix \ref{appenEk}. Thus if we prove that $V_{N,d,b,g}$ is simple semi-homogeneous, we prove the expected isomorphism up to a flat twist.

Here we only outline the proof of simple and semi-homogeneity of $V_{N,d,b,g}$, since the full details are beyond the scope of this paper. Let $X=\mathrm{Sym}^N(\Sigma), A=\mathrm{Pic}^N(\Sigma), B=\mathrm{Pic}^d(\Sigma)$. For $N$ large enough, the maps given by Abel--Jacobi map  $\sigma:X\rightarrow A, \rho:=\sigma\times \operatorname{id}:X\times B\rightarrow A\times B$ have projective bundle structure \cite{mattuck1961symmetric}. Let $\theta_A, \theta_B, \eta_{A,B}$ be theta and "mixed" classes in \cite[Section 2.1]{klevtsov2025chern}. Choose identification $B\simeq \widehat{A}$ and normalized Poincar\'e bundle $\mathcal{P}$ on $A\times B$ with $c_1(\mathcal{P})=\eta_{A,B}$. Using projective bundle structure $\sigma: X\simeq \mathbb{P}(E)\rightarrow A$ and Chern class computation of \cite{klevtsov2025chern}, one can show that $\mathcal{M}:=\rho_*(\mathbf{L}_b)$ is a line bundle on $A\times B$ with $\mathbf{L}_b \simeq \rho^*\mathcal{M}, c_1(\mathcal{M})=b\theta_A-\eta_{A,B}$. Let $\mathcal{T}_A$ be a theta line bundle on $A$ with $c_1(\mathcal{T}_A)=\theta_A$. Then by the K\"unneth formula, there are $\alpha\in \widehat{A}, \beta\in \widehat{B}$ such that $\mathcal{M}\simeq \operatorname{pr}_A^*(\mathcal{T}_A^b \otimes \alpha)\otimes \mathcal{P}^{-1}\otimes \operatorname{pr}_B^*\beta$. Since $V_{N,d,b,g}:=(\operatorname{pr}_B)_* \rho_* \mathbf{L}_b \simeq (\operatorname{pr}_B)_* \mathcal{M}$, this realizes $V_{N,d,b,g}$ as a Fourier--Mukai transform of an ample line bundle $L_A:=\mathcal{T}_A^b \otimes \alpha$ ($L_A$ is ample and satisfies I.T. with $i(L_A)=0$). Then \cite[Corollary 2.5, Proposition 3.11 (1)]{mukai1981duality} and \cite[Proposition 7.3]{mukai1978semi} show that $V_{N,d,b,g}\simeq \widehat{L_A}\otimes \beta$ is simple semi-homogeneous.

\subsection{Asymptotic projective flatness of the physical \texorpdfstring{$L^2$}{L2} metric}\label{subsectionproof}

Now we compare the (dual of) the theta metric on $\mathcal{V}_k$ considered in Section \ref{ABsubsection} and the physical $L^2$ inner product of states in \eqref{klevformula}. Throughout this subsection, we fix $k, g>1$ and let $M=s=0$, and consider $N\rightarrow \infty$. The main result of this subsection is to show that the physical $L^2$ metric is asymptotically identical to the natural theta metric, which implies that the Hall conductance result of the Section \ref{ABsubsection} is asymptotically true even before taking the average over $Jac(\Sigma)$. The setting and the proof are quite complicated, so we divide the following discussions into several separate paragraphs.

\paragraph{Setting and the main result}

To define a physical $L^2$ metric on Laughlin states \ref{klevformula}, we need to specify the hermitian metrics on the surface $\Sigma$ and the magnetic line bundle taking account of AB flux. We will follow the metric choice of \cite[Section 4.7]{klevtsov2019laughlin}. 

On the Riemann surface $\Sigma$ , we choose the Arakelov metric $g_a$, and denote the canonical metric by $g_c$ (see \cite[Section 4.6]{klevtsov2019laughlin}). To specify the hermitian metric on the magnetic bundle, let $L_\xi\simeq \mathcal{O}(D_\xi)$ be the magnetic line bundle corresponding to $\xi\in Jac(\Sigma)$ \eqref{jaccoor}. More precisely, fix a degree $N_\Phi$ line bundle $L_0$ on $\Sigma$ and a connection $A_0$ on it. For later convenience, choose $L_0\simeq \mathcal{O}(D_0)$ so that 
\begin{equation}\label{choice}
    \mathcal{I}(D_0)=-\frac{N_\Phi}{g-1}\Delta.
\end{equation}
For any other degree $N_\Phi$ line bundle $L_\xi$ with connection $A$, $L_\xi\otimes L_0^{-1}\simeq \mathcal{O}(D)$ is a topologically trivial line bundle with flat connection $A_\xi$. Hence there exists a smooth global section $e$. Let $s=fe$ with $(s)=D=D_\xi-D_0$ and
\begin{equation*}
    \eta:=\frac{\nabla s}{s}=\frac{df}{f}+A_\xi
\end{equation*}
be a global meromorphic 1-form on $\Sigma$. If we denote $D$ by $D=\sum n_P P$, $\eta$ has poles at $P$ with residue $n_P$. Note that by the Riemann bilinear relation,
\begin{equation*}
   2\pi i \, \xi_i=\int_\Sigma A_\xi^{0,1}\wedge \omega_i = \tau_{ij} \int_{a_j} A_\xi - \int_{b_i} A_\xi \equiv \tau_{ij} \int_{a_j} \eta - \int_{b_i} \eta \quad(\operatorname{mod} \, 2\pi i\Lambda). 
\end{equation*}
Consider $\eta_D:=\sum n_P \tau_{PQ}$ where $\tau_{PQ}$ is the normalized differential of the third kind in \cite[(3.6.1)]{farkas1992riemann}. Then $\eta-\eta_D$ is holomorphic and can be expanded in terms of $\omega$:
\begin{equation*}
    \eta-\eta_D=\left( \int_{a_j}\eta\right) \omega_j.
\end{equation*}
By \cite[(3.6.2)]{farkas1992riemann}, this implies
\begin{equation*}
    \int_{b_i}\eta -\tau_{ij}\int_{a_j}\eta = 2\pi i\, \mathcal{I}(D)_i.
\end{equation*}
As a result, we obtain under the choice \eqref{choice},
\begin{equation}\label{xixi}
    \mathcal{I}(D_\xi)=\mathcal{I}(D+D_0)=-\frac{N_\Phi}{g-1}\Delta-\xi.
\end{equation}
Let $\Delta^z:=\Delta+(g-1)\mathcal{I}(z)$ and define
\begin{equation*}
    \mathcal{A}_N(z_j):=\sum_{j=1}^N \mathcal{I}(z_j)+\frac{N}{g-1}\Delta=\sum_{j=1}^N \frac{\Delta^{z_j}}{g-1},
\end{equation*}
so that the Laughlin states \eqref{klevformula} takes the form
\begin{equation}\label{klevformula2}
    F_{r,N}(z_j;\xi)=\Theta^{(k)}_r\left(\mathcal{A}_N(z_j)+\frac{\xi}{k}\,\middle|\, \tau\right)\prod_{i<j}E(z_i,z_j)^k \prod_{j} \sigma(z_j)^{k(N-1)/g}.
\end{equation}
By \eqref{xixi}, the hermitian metric \cite[(3.26)]{klevtsov2019laughlin} on $L_\xi$ now takes the simple form
\begin{equation}\label{hxi}
    \log h_\xi(z)=-\frac{2\pi N_\Phi}{g(g-1
    )^2}|\operatorname{Im}\Delta^z|^2_{\tau}-\frac{4\pi}{g-1}\langle \operatorname{Im}\Delta^z, \operatorname{Im}\xi \rangle_\tau
\end{equation}
where we denote by $\langle x,y\rangle_\tau$ the inner product $x^T(\operatorname{Im}\tau)^{-1}y$ on $\mathbb{R}^g$. The physical Gram matrix is then given by
\begin{equation*}
    (\mathbf{G}_N)_{rs}(\xi)=\int_{\Sigma^N} \overline{F_{r,N}(z_j;\xi)}F_{s,N}(z_j;\xi)\prod_{j}^N h_{\xi}(z_j) dV_{g_a}(z_j).
\end{equation*}
This defines a hermitian metric on $\mathcal{V}_{k}$ and its Chern connection $\nabla_{N}$ with curvature $F_{\nabla_{N}}$. 

Let 
\begin{equation*}
    H(\xi)=\exp \left( \frac{2\pi}{k}|\operatorname{Im}\xi|^2_\tau \right)\mathbf{1}_{k^g}
\end{equation*}
be the dual theta metric on $\mathcal{V}_{k}$, with corresponding Chern connection $\nabla$. Although we are using the same symbol $H$ as \eqref{H}, here it denotes the dual of it. By \eqref{curvature}, we have
\begin{equation*}
    F_{\nabla}=\frac{2\pi i}{k}\omega_{\mathrm{flux}}\mathbf{1}_{k^g}.
\end{equation*}
Our main theorem is the following.
\begin{theorem}\label{thm}
There are (explicit) positive constants $C_N$ such that 
\begin{equation}\label{thmlimit}
   C_N^{-1}H^{-1}(\xi)\mathbf{G}_N(\xi) \longrightarrow \mathbf{1}_{k^g}
\end{equation}
in $C^2$ as $N\rightarrow\infty$. This implies
\begin{equation*}
    F_{\nabla_N}=\frac{2\pi i}{k}\omega_{\mathrm{flux}}\mathbf{1}_{k^g}+o(1).
\end{equation*}
That is, the physical Berry curvature is asymptotically projectively flat.
\end{theorem}

The proof goes as follows. First, we will factorize $H^{-1}\mathbf{G}_N$ into theta part and $\xi$-independent Coulomb part. We then show that the pushforward of the Coulomb part converges weakly to Haar measure on the Jacobian. The Haar average of the theta factor is then a scalar multiple of the identity. The essential ingredient of the proof is the construction of the "asymptotic" magnetic translations on $\Sigma$ for $g>1$.

\paragraph{Factorization of the Gram matrix into theta and Coulomb parts}

Define the matrix-valued function on $Jac(\Sigma)$
\begin{equation*}
    (\mathsf{T}_k)_{rs}(u):=e^{-2\pi k|\operatorname{Im}u|^2_\tau} \overline{\Theta^{(k)}_r(u \mid \tau)}\Theta^{(k)}_s(u \mid \tau).
\end{equation*}
This will be the theta part of $H^{-1}\mathbf{G}_N$. For the Coulomb part, let $\mu$ be the normalized volume form of $g_c$, and $G$ its scalar Green function with normalization
\begin{equation*}
    G(z,w)=-\log d(z,w)+O(1),\quad \int_\Sigma G(z,w) d\mu(w)=0. 
\end{equation*}
In the notation of \cite{klevtsov2019laughlin}, $G=G_{g_c}=G^A_{g_a}$. Write $dV_{g_a}=\varrho \,\mu$ with a strictly positive function $\varrho$. Define the Coulomb law on $\Sigma^N$ and its pushforward under the Abel--Jacobi map by
\begin{equation}\label{Claw}
    d\mathbb{P}_N(z_j):=\frac{1}{\mathcal{Z}_N}e^{-k\sum_{i\neq j}G(z_i,z_j)}\prod_{j=1}^N dV_{g_a}(z_j), \qquad \nu_N:=(\mathcal{A}_N)_*\mathbb{P}_N.
\end{equation}

\begin{proposition}\label{propofac}
There are constants $B_N>0$ such that
\begin{equation}\label{factor}
    H(\xi)^{-1}\mathbf{G}_N(\xi)=B_N\int_{Jac(\Sigma)} \mathsf{T}_k\left( a+\frac{\xi}{k}\right) d\nu_N(a).
\end{equation}
Let $m$ be a normalized Haar measure on $Jac(\Sigma)$. Then for every $b\in Jac(\Sigma)$,
\begin{equation}\label{Haar}
    \int_{Jac(\Sigma)} \mathsf{T}_k\left( a+b\right) dm(a) =c_k \,\mathbf{1}_{k^g}, \quad c_k=\frac{1}{\sqrt{\det (2k\operatorname{Im}\tau)}}.
\end{equation}
\end{proposition}

\begin{proof}
The non-theta part of $\mathbf{G}_N$ is given by
\begin{equation}\label{star}
    \int_{\Sigma^N} \prod_{i<j}|E(z_i,z_j)|^{2k} \prod_j |\sigma(z_j)|^{2k(N-1)/g} h_\xi(z_j) \,dV_{g_a}(z_j).
\end{equation}
By \cite[(4.42)-(4.43)]{klevtsov2019laughlin}, we have
\begin{align*}
    \log |E(z,w)|^2&=-2G(z,w)-\frac{1}{2}\log g_a(z)-\frac{1}{2}\log g_a(w)+2\pi|\operatorname{Im}\mathcal{I}(z-w)|^2_\tau,\\
    \log g_a(z) &= 2\log c_\Sigma +\frac{2}{g}\log|\sigma(z)|^2+\frac{4\pi}{g(g-1)}|\operatorname{Im}\Delta^z|^2_\tau.
\end{align*}
Take the logarithm of the integrand in \eqref{star}. The first term coming from prime forms is then
\begin{align*}
    &-k\sum_{i\neq j}G(z_i,z_j)+2\pi k\sum_{i<j}|\operatorname{Im}\mathcal{I}(z_i-z_j)|^2_\tau-kN(N-1)\log c_\Sigma\\
    &\qquad -\frac{k(N-1)}{g}\sum_j \log |\sigma(z_j)|^2 -\frac{2\pi k(N-1)}{g(g-1)}\sum_j |\operatorname{Im}\Delta^{z_j}|^2_\tau.
\end{align*}
The last two terms cancel out terms from $\prod|\sigma|^{2k(N-1)/g}h_\xi$ (recall \eqref{hxi}) and we are left with
\begin{align}
    &-k\sum_{i\neq j}G(z_i,z_j)+2\pi k\sum_{i<j}|\operatorname{Im}\mathcal{I}(z_i-z_j)|^2_\tau-kN(N-1)\log c_\Sigma \nonumber\\
    &\qquad -\frac{2\pi kN}{(g-1)^2}\sum_j |\operatorname{Im}\Delta^{z_j}|^2_\tau -\frac{4\pi}{g-1}\sum_j\langle \operatorname{Im}\Delta^{z_j}, \operatorname{Im}\xi\rangle_\tau.\label{222}
\end{align}
Here we used the relation \eqref{Sneutrality}. Now let $v_j:=\operatorname{Im}\Delta^{z_j}/(g-1)$ and $a:=\sum_j v_j$. Then using $\operatorname{Im}\mathcal{I}(z_i-z_j)=v_i-v_j$ and $\sum_{i<j}|v_i-v_j|_\tau^2=N\sum_j |v_j|^2_\tau-|\sum_j v_j|^2_\tau$, we can collect the second and fourth terms in \eqref{222} as $-2\pi k |a|^2_\tau$. Also, apply $\sum_j \operatorname{Im}\Delta^{z_j}=(g-1)a$ to the last term in \eqref{222} and complete the square. The end result is
\begin{equation*}
    -k\sum_{i\neq j}G(z_i,z_j)-2\pi k\left|\operatorname{Im}\left(\mathcal{A}_N(z_j)+\frac{\xi}{k}\right)\right|^2_\tau+\frac{2\pi}{k}\left|\operatorname{Im}\xi \right|^2_\tau-kN(N-1)\log c_\Sigma
\end{equation*}
using the fact that $a=\operatorname{Im}\mathcal{A}_N$. Note that we express it in the same variable as the theta argument in \eqref{klevformula2}. The formula \eqref{factor} now follows with $B_N=c_\Sigma^{-kN(N-1)}\mathcal{Z}_N$.

Writing $dm=dx\wedge dy, a=x+\tau y$ with $x,y\in[0,1]^g$, the formula \eqref{Haar} follows from direct calculation of Gaussian integrals.
\end{proof}

\paragraph{Convergence of the Coulomb part}

Here we prove the following result.
\begin{proposition}\label{magtransl}
    The pushforward of the Coulomb law under the Abel--Jacobi map $\nu_N$ converges weakly to Haar measure $m$ as $N\rightarrow\infty$. 
\end{proposition}
To prove this, we need two lemmas.
\begin{lemma}\label{empirical}
Under $\mathbb{P}_N$, the empirical measure $L_N:=N^{-1}\sum_j \delta_{z_j}$ converges weakly to $\mu$ in probability.
\end{lemma}

\begin{proof}
Define the law $\mathbb{Q}_N$ as in \eqref{Claw} with $dV_{g_a}$ replaced by $d\mu$. Then $\mathbb{Q}_N$ is the Gibbs law with total energy $N^{-2}\sum_{i<j}G(z_i,z_j)/2\pi$ and inverse temperature $\beta_N=4\pi kN^2$ in \cite{garciazelada2019concentration}. By \cite[Theorem 1.1]{garciazelada2019concentration},
\begin{equation*}
    \mathbb{Q}_N(W_1(L_N,\mu)\geq r) \leq \exp\left(-\pi N^2 r^2+\frac{k}{2}N\log N +C N \right)
\end{equation*}
for some constant $C$, where $W_1$ is the Wasserstein distance with respect to $g_c$. Recall that $dV_{g_a}=\varrho\,\mu$. Choose positive constants $\varrho_{-},\varrho_+$ such that $0<\varrho_{-}<\varrho<\varrho_+$. Then for any event $E$,
\begin{equation*}
    \mathbb{P}_N(E)=\frac{\mathbb{E}_{\mathbb{Q}_N} \left[\mathbf{1}_E \prod_j \varrho(z_j)\right]}{\mathbb{E}_{\mathbb{Q}_N} \left[\prod_j \varrho(z_j)\right]} \leq \left(\frac{\varrho_+}{\varrho_{-}}\right)^N \mathbb{Q}_N(E).
\end{equation*}
Letting $E=\{ W_1(L_N, \mu)\geq r\}$, we get $\mathbb{P}_N(W_1(L_N,\mu)\geq r)\rightarrow 0$. Since $W_1$ metrizes weak convergence, we prove the lemma.
\end{proof}

\begin{lemma}\label{transl}
Let $\Phi_t$ be the flow of a smooth vector field on $\Sigma$ preserving $\mu$. Let $T_t:=\prod_j^N \Phi_t$ act on $\Sigma^N$. Then
\begin{equation*}
    \|(T_{1/N})_*\mathbb{P}_N-\mathbb{P}_N\|_{\mathrm{TV}} \longrightarrow 0.
\end{equation*}
\end{lemma}

\begin{proof}
Denote the density function of the measure $\mathbb{P}_N$ with respect to $\mu^{\otimes N}$ by $p_N$:
\begin{equation*}
    d\mathbb{P}_N=p_N\, d\mu^{\otimes N}, \quad p_N=\mathcal{Z}_N^{-1}e^{-\mathcal{H}_N}
\end{equation*}
    with
\begin{equation*}
    \mathcal{H}_N(z_j)=k\sum_{i\neq j}G(z_i,z_j)+\sum_j U(z_j), \quad U=-\log \varrho.
\end{equation*}
Since $T_t$ preserves $d\mu^{\otimes N}$, we have
\begin{equation*}
    d(T_t)_*\mathbb{P}_N=p_t\, d\mu^{\otimes N}, \quad p_t=p_N(T_{-t}(z_j)).
\end{equation*}
To estimate the total variation, which is half of the $L^1$-distance between the densities, we will estimate $\|p_t-p_N\|_{L^1(\mu^N)}$ for the rest of the proof.

Let $X$ be a $\mu$-preserving vector field on $\Sigma$ and $\Phi_t$ be its flow. Let $X_N:=\sum_{j=1}^N X_j$ be the generator of $T_t$ (with $X_j$ acting on the $j^{th}$ copy of $\Sigma$).
Consider 
\begin{equation*}
    \mathcal{Z}_N(t):=\int_{\Sigma^N}e^{-\mathcal{H}_N(T_t(z_j))}d\mu^{\otimes N}.
\end{equation*}
It is constant in $t$. To differentiate it twice with respect to $t$, note that
\begin{equation*}
    \left.\frac{d^2}{dt^2}\right|_{t=0} e^{-\mathcal{H}_N(T_t(z_j))}=\left[ (X_N\mathcal{H}_N)^2-X_N^2\mathcal{H}_N \right]e^{-\mathcal{H}_N}.
\end{equation*}
Thus $\mathcal{Z}_N''(0)/\mathcal{Z}_N(0)=0$ implies
\begin{equation}\label{IN}
    I_N:=\mathbb{E}_{\mathbb{P}_N}(X_N\mathcal{H}_N)^2=\mathbb{E}_{\mathbb{P}_N} (X^2_N\mathcal{H}_N).
\end{equation}
   
Since $\partial_t p_t=((X_N\mathcal{H}_N)\circ T_{-t} )p_t$, we have $\|\partial_t p_t\|_{L^1(\mu^N)} =\mathbb{E}_{\mathbb{P}_N} |X_N\mathcal{H}_N|\leq \sqrt{I_N}$ by the Cauchy--Schwarz inequality. That is,
\begin{equation}\label{final}
    \|(T_{t})_*\mathbb{P}_N-\mathbb{P}_N\|_{\mathrm{TV}}=\frac{1}{2}\|p_t-p_N\|_{L^1(\mu^N)}\leq \frac{|t|}{2}\|\partial_t p_t\|_{L^1(\mu^N)}\leq  \frac{|t|}{2}\sqrt{I_N}.
\end{equation}
Hence if we can show that $I_N/N^2\rightarrow 0$, we are done.

Define 
\begin{equation*}
    G^{(1)}(z,w):=(X_z+X_w)G(z,w), \quad G^{(2)}(z,w):=(X_z+X_w)^2G(z,w).
\end{equation*}
Since locally $G(z,w)\sim -\log|z-w|$, $G^{(1)}$ and $G^{(2)}$ are bounded on $\Sigma\times\Sigma$ and continuous off the diagonal. To see this, for example, locally write $X_z=X(z)\cdot \nabla_z, X_w=X(w)\cdot \nabla_w$ and check that
\begin{equation*}
(X_z+X_w)(-\log|z-w|)=-\frac{(z-w)\cdot (X(z)-X(w))}{|z-w|^2},   
\end{equation*}
which is $O(1)$ because $X$ is differentiable and hence $X(z)-X(w)=O(|z-w|)$. This shows that $G^{(1)}$ is bounded and $G^{(2)}$ can be treated similarly. Extend $G^{(2)}$ by zero on diagonal. Note that since $\Phi_t$ preserves $\mu$, 
\begin{equation*}
    \int_{\Sigma^2}G(\Phi_t z,\Phi_t w)d\mu(z)d\mu(w)=\int_{\Sigma^2}G(z,w)d\mu(z)d\mu(w).
\end{equation*}
Taking differentiation in $t$ twice, we get
\begin{equation}\label{G2zero}
    \int_{\Sigma^2}G^{(2)}(z,w)d\mu(z)d\mu(w)=0.
\end{equation}

Now we estimate the right hand side of \eqref{IN}. First, note that
\begin{equation}\label{INR}
    X_N^2\mathcal{H}_N=k\sum_{i\neq j}G^{(2)}(z_i,z_j)+\sum_j X^2U(z_j).
\end{equation}
Using the empirical measure $L_N=N^{-1}\sum_j \delta_{z_j}$, we have
\begin{equation*}
    \frac{1}{N^2}\sum_{i\neq j}G^{(2)}(z_i,z_j)=\int_{\Sigma^2}G^{(2)} dL_N^{\otimes 2}.
\end{equation*}
By Lemma \ref{empirical}, under $\mathbb{P}_N$, $L_N^{\otimes 2}$ weakly converges to $\mu^{\otimes 2}$ in probability. Since $G^{(2)}$ is bounded and continuous outside of diagonal which has measure zero with respect to $\mu^{\otimes 2}$, by the Portmanteau theorem and \eqref{G2zero}, we obtain 
\begin{equation*}
    \frac{1}{N^2}\sum_{i\neq j}G^{(2)}(z_i,z_j)\rightarrow \int_{\Sigma^2}G^{(2)}(z,w)d\mu(z)d\mu(w)=0
\end{equation*}
in probability. Moreover, since $|\int_{\Sigma^2}G^{(2)} dL_N^{\otimes 2}|\leq \|G^{(2)}\|_\infty$, the convergence is in $L^1(\mathbb{P}_N)$. That is, we get
\begin{equation}\label{est1}
    \mathbb{E}_{\mathbb{P}_N}\frac{1}{N^2}\sum_{i\neq j}G^{(2)}(z_i,z_j)\longrightarrow 0.
\end{equation}
For the second term of \eqref{INR}, since $X$ and $U$ are smooth, we have
\begin{equation}\label{est2}
    \left| \sum_j X^2 U(z_j)\right|\leq N\|X^2 U\|_\infty=O(N).
\end{equation}
Combining \eqref{est1} and \eqref{est2}, we finally obtain
\begin{equation*}
    \frac{I_N}{N^2}=k\mathbb{E}_{\mathbb{P}_N}\frac{1}{N^2}\sum_{i\neq j}G^{(2)}(z_i,z_j)+O(N^{-1})\longrightarrow 0.
\end{equation*}
By \eqref{final}, this completes the proof.
\end{proof}

In genus $g=1$ case, the existence of magnetic translations plays the key role in the study of the $L^2$ metric of Laughlin states \cite{Burban:2023cii}. On higher genus surfaces, no such translations exist, but the following proof constructs the analogous diffeomorphisms on $\Sigma$ which asymptotically recover translations on $Jac(\Sigma)$ under Abel--Jacobi map.

\begin{proof}[Proof of Proposition \ref{magtransl}.]
We start with the isomorphism
\begin{equation*}
    H^1(\Sigma;\mathbb{R})\longrightarrow T_0 Jac(\Sigma), \quad [b]\longmapsto \int_\Sigma d\mathcal{I}\wedge b.
\end{equation*}
Fix $\eta\in T_0 Jac(\Sigma)$. Choose a smooth closed form $b$ mapping to $\eta$, and define $X$ by $\iota_X \mu=b$. Cartan's formula gives,
\begin{equation}\label{cartan}
    \mathcal{L}_X \mu=db=0,\quad \int_\Sigma d\mathcal{I}(X)d\mu=\int_\Sigma d\mathcal{I}\wedge b=\eta.
\end{equation}
Let $\Phi_t, T_t$ be as in Lemma \ref{transl}. By Taylor expansion of Abel--Jacobi map $\mathcal{A}_N$, Lemma \ref{empirical} and \eqref{cartan}, we have
\begin{equation}\label{11111}
    \mathcal{A}_N(T_{1/N}(z_j))- \mathcal{A}_N(z_j)=\frac{1}{N}\sum_j d\mathcal{I}(X)(z_j)+O(N^{-1})\longrightarrow \eta
\end{equation}
in probability. This is uniform in $z_j$. Also, for $f\in C(Jac(\Sigma))$, Lemma \ref{transl} implies
\begin{equation}\label{2222}
    \mathbb{E}_{\mathbb{P}_N}f(\mathcal{A}_N(T_{1/N}(z_j)))-\mathbb{E}_{\mathbb{P}_N}f(\mathcal{A}_N(z_j))\longrightarrow 0.
\end{equation}
Combining \eqref{11111} and \eqref{2222}, we obtain
\begin{equation*}
    \mathbb{E}_{\mathbb{P}_N}f(\mathcal{A}_N(z_j)+\eta)-\mathbb{E}_{\mathbb{P}_N}f(\mathcal{A}_N(z_j))\longrightarrow 0.
\end{equation*}
Thus every subsequential limit of $\nu_N$ is translation invariant. Since $Jac(\Sigma)$ is compact, this implies that $\nu_N$ converges to Haar measure $m$.
\end{proof}

\paragraph{Proof of the main result}
Now the main result follows immediately.
\begin{proof}[Proof of Theorem \ref{thm}]
Set $C_N=c_k \,B_N$. By Proposition \ref{propofac}, 
\begin{equation*}
    C_N^{-1}H^{-1}(\xi)\mathbf{G}_N(\xi)-\mathbf{1}_{k^g}=c_k^{-1}\int_{Jac(\Sigma)} \mathsf{T}_k \left(a+\frac{\xi}{k}\right)d(\nu_N-m)(a).
\end{equation*}
Differentiating with respect to $\xi$ and applying Proposition \ref{magtransl}, we prove \eqref{thmlimit}.
\end{proof}
Here our result concerns AB fluxes $\xi$. We believe that similar analysis can be done for quasi-hole positions.

\section{Conclusion}

In this paper we revisited the explicit construction of higher genus Laughlin states in the spirit of \cite{moore1991nonabelions}. We then studied the explicit dependence on quasi-hole positions and AB fluxes. For quasi-hole transport, the local and global monodromies reproduce the familiar fractional statistics, fractional AB phases and finite Heisenberg algebra \eqref{algebra}. For AB fluxes, the topology of the bundle and the Niu--Thouless--Wu formula give the averaged Hall conductance $1/k$. This cohomological result is independent of the choice of hermitian metric.

We also reproduced the Wen--Zee shift formula by including the Wen--Zee coupling and gravitational spin on the CS side. Setting the gravitational spin to $s=0$, we recover the explicit Laughlin states of \cite{klevtsov2019laughlin} from the higher genus Moore--Read construction. Moreover, we constructed the bundle of Laughlin states $\mathcal{V}_k$ over $Jac(\Sigma)$ with a projectively flat reference Chern connection and computed its characteristic classes. These coincide with those of the direct image Laughlin bundle of \cite{klevtsov2025chern}, and the bundles are holomorphically isomorphic up to a flat twist. We then proved that the physical $L^2$ metric is in fact asymptotically isometric to the reference theta metric, thereby showing that the physical Berry curvature is indeed projectively flat, at least asymptotically.

Finally, we would like to mention a possible connection with complex geometry. For example, it might be interesting to compare with Berman's canonical random point processes in K\"ahler geometry \cite{Berman:2010kq}. These processes are obtained by normalizing a determinant-type many-particle density constructed from pluricanonical sections, and the resulting normalization integral is interpreted as a partition function. There is a conjecture concerning how this partition function behaves in the large particle number limit \cite{Berman:2023jcs}. There is also a more direct generalization of the integer quantum Hall effect for higher dimensions \cite{Berman:2014aya, Eum:2025zxg}. The physical $L^2$ Gram matrix of the Laughlin states may be regarded as an interacting, matrix-valued analogue of such partition functions. It would therefore be interesting to see how this can be generalized to higher dimensions, and whether its asymptotics admit a similar interpretation in complex geometry.

\appendix
\section{Gauge and frame conventions}\label{appen}

Throughout this paper we used the convention $\nabla=d+A$ except in the CS action. Under the physics convention (which we used in the CS action) $\nabla=d-iA_{\mathrm{ph}}$, we get $F_{\mathrm{ph}}=dA_{\mathrm{ph}}=iF$ and
\begin{equation*}
    c_1(L)=\left[ \frac{iF}{2\pi}\right]=\left[ \frac{F_{\mathrm{ph}}}{2\pi}\right].
\end{equation*}
For example, the symmetric gauge used in \eqref{Laughlin} is
\begin{align*}
    A&=-\frac{\bar{z}}{4}\,dz+\frac{z}{4}\,d\bar{z} =\frac{i}{2}y\,dx- \frac{i}{2}x\,dy,\\
    A_{\mathrm{ph}}&=iA=-\frac12 y\,dx+\frac12 x\,dy.
\end{align*}

Let $X$ be a complex parameter space and let $\ket{\psi(\lambda)}$ be a local holomorphic frame. For simplicity, assume the dimension of the Hilbert space is $1$. Denote the norm by
\begin{equation*}
    Z(\lambda,\bar\lambda)=\braket{\psi(\lambda)|\psi(\lambda)}.
\end{equation*}
Under this holomorphic frame, the Chern connection is given by (see \cite{kobayashi2014differential})
\begin{equation*}
    \nabla = d +A_{\mathrm{hol}},\qquad A_{\mathrm{hol}}=\partial \log Z.
\end{equation*}
Let $\ket{\widehat{\psi}}=Z^{-1/2}\ket{\psi}$ be the normalized frame. This is no longer holomorphic in general. Under this unitary frame, the Chern connection is given by
\begin{equation*}
    \nabla = d +A_{u},\qquad A_{u}=A_{\mathrm{hol}}+d\log Z^{-1/2}=\frac{1}{2}\partial \log Z-\frac{1}{2}\bar\partial \log Z.
\end{equation*}
This explains \eqref{Berry}, \eqref{Berryanti} and \eqref{diffeqn1}.

\section{The automorphy bundle $\mathcal E_k$}\label{appenEk}

The factor of automorphy $J_{n,m}$ in \eqref{factorofaut} is written explicitly as
\begin{equation*}
 \left(J_{n,m}(\xi)\right)_{\alpha,\beta}=\exp\left(\frac{2\pi i}{k}\alpha^T n -\frac{\pi i}{k}m^T\tau m -\frac{2\pi i}{k}m^T\xi \right) \delta_{\alpha+m,\beta}.
\end{equation*}
We decompose it into a unitary piece and a non-unitary piece to make computation easier. Let
\begin{equation*}
q_m(\xi)= \exp\left( -\frac{\pi i}{k}m^T \tau m -\frac{2\pi i}{k}m^T\xi \right),\qquad (U_{n,m}v)_r=e^{2\pi i r^Tn/k}v_{r+m},
\end{equation*}
so that $J_{n,m}(\xi)=q_m(\xi)U_{n,m}$. Write
\begin{equation*}
    H(\xi)=h(\xi)\mathbf{1}_{k^g}, \quad h(\xi)=\exp\left(-\frac{2\pi}{k}y^T Y^{-1}y\right), \quad Y:=\mathrm{Im}\,\tau, \quad  y:=\mathrm{Im}\,\xi.
\end{equation*}
Then 
\begin{equation*}
h(\xi+n+\tau m)=h(\xi)\exp\left( -\frac{2\pi}{k}m^T Ym -\frac{4\pi}{k}m^T y\right).
\end{equation*}
Since $U_{n,m}$ is unitary and
\begin{equation*}
    |q_m(\xi)|^2 =\exp\left( \frac{2\pi}{k}m^TYm +\frac{4\pi}{k}m^T y \right),
\end{equation*}
we get 
\begin{equation}\label{metricproof}
  J_{n,m}(\xi)^\dagger H(\xi+n+\tau m) J_{n,m}(\xi)=H(\xi).
\end{equation}
Thus $H$ descends to a Hermitian metric on
$\mathcal{E}_k$.

Now we show that $\mathcal{E}_k$ is simple semi-homogeneous. Semi-homogeneity is immediate since
\begin{equation}\label{semihom}
    J_{n,m}(\xi+\xi')=\exp \left( -\frac{2\pi i}{k}m^T \xi'\right) J_{n,m}(\xi).
\end{equation}
It also follows from projective flatness. To show simpleness, let $f(\xi)$ be a holomorphic endomorphism of $\mathcal{E}_k$. Then
\begin{equation}\label{Shur}
    f(\xi+n+\tau m)=J_{n,m}(\xi)f(\xi)J_{n,m}(\xi)^{-1}=U_{n,m}f(\xi)U_{n,m}^{-1}.
\end{equation}
Since $U_{n,m}$ are unitary, $||f(\xi)||_{op}$ is $\Lambda$-periodic. By Liouville theorem, this implies that $f(\xi)$ is constant. Letting $n=0, m=0$ in turn in \eqref{Shur}, some tedious algebra gives $f=C\cdot\mathbf{1}_{k^g}$ for some constant $C$.

\acknowledgments
The author would like to thank Semyon Klevtsov for helpful discussions on Laughlin states. This work was supported by the National Research Foundation of Korea (NRF) grant funded by the Korea government (MSIT) RS-2024-00346651.
\paragraph{Use of generative AI.}
The author used ChatGPT 5.6 sol and 6 Astra to assist with language editing and mathematical content of Section \ref{subsectionproof} and the final paragraph of Section \ref{4.2}. All AI-assisted output was independently checked and revised by the author, who takes full responsibility.

\bibliographystyle{JHEP}
\bibliography{biblio.bib}

\end{document}